\documentclass[prl,aps,twocolumn,showpacs]{revtex4-1}
\usepackage{graphicx}
\usepackage{wasysym}
\usepackage{bm}
\usepackage{times}
\usepackage{color}
\usepackage[colorlinks=true, urlcolor=blue, linkcolor=red, citecolor=blue, pdftex]{hyperref}

\newcommand{\beq}{\begin{equation}}
\newcommand{\eeq}{\end{equation}}
\newcommand{\bea}{\begin{eqnarray}}
\newcommand{\eea}{\end{eqnarray}}

\def\bei{\begin{itemize}}
\def\eei{\end{itemize}}

\def\unue#1{{\it #1:}}

\begin{document}

\title{Quantum simulations made easy plane}

\author{Andreas M. L\"auchli}
\affiliation{Institute for Theoretical Physics, University of 
Innsbruck, A-6020 Innsbruck, Austria}
\author{R. Moessner}
\affiliation{Max Planck Institut f\"ur Physik komplexer Systeme, D-01187 Dresden, Germany}

\date{\today}

\begin{abstract}
Ever since Heisenberg's proposal of a quantum-mechanical origin  of ferromagnetism in 1928, 
the spin model named after him has been central to advances in
magnetism, featuring in proposals of novel many-body states such as antiferromagnets,
emergent gauge fields in their confined (valence bond crystal) and deconfined 
(resonating valence bond spin liquids) versions.  Between them, these cover
much of our understanding of modern magnetism specifically and topological states of matter in general. 
Many  exciting phenomena 
predicted theoretically still await experimental realisation, and 
cold atomic 
systems hold the promise of acting as analogue
'quantum simulators' of the relevant theoretical models, for which  
ingenious and intricate
set-ups have been proposed. Here, we identify a new class of {\em
particularly simple quantum simulators} exhibiting many such phenomena but 
obviating the need for fine-tuning and for
amplifying perturbatively weak superexchange 
or  longer-range interactions. Instead they require only moderate {\em
on-site} interactions on top of uncorrelated, {\it one-body} hopping--ingredients 
 already available with present experimental technology.  Between them, they 
realise some of the most interesting phenomena, such as 
 emergent synthetic gauge
fields, resonating valence bond phases, and even the celebrated yet
enigmatic spin liquid phase of the kagome lattice. 
\end{abstract}


\maketitle
Quantum spin liquids and other exotic magnetic states are among the theoretically most
interesting yet experimentally elusive instance of collective quantum phenomena, underpinning
as they do the counterintuitive emergence of new particles from the `decay' 
of electrons in a magnet \cite{Anderson1987}, 
such as holons and spinons which carry the electric and magnetic properties
of an electron independently, or the entirely novel Majorana particles underpinning
attemps to build so-called topological quantum computers protected against decoherence
due to noise from the environment \cite{Kitaev2003}.  

Some of the earliest proposals for quantum simulators have thus aimed at generating
such exotic spin physics, e.g.\ in the case of the Kitaev model \cite{Kitaev2006} with its topological 
spin liquid phase 
 \cite{Duan2003}. 
The challenges facing such a programme are manifold and formidable. 
Firstly, it is necessary to generate the requisite degrees of freedom; secondly, they need to inhabit
a suitable optical lattice; thirdly, their interactions must be nontrivial and 
satisfy symmetries of the target Hamiltonian; and finally, the system in question must be cooled to
a  temperature below which the target many-body states are stable. All these points have seen 
much progress on the theoretical front over the years, and considerable
effort has been invested in identifying promising settings in which naturally occurring strong interactions
-- be it e.g.\ of dipolar nature or through Rydberg physics -- can lead to interesting many-body states.

Here we identify a class of particularly
simple spin one-half model Hamiltonians which go along
with particularly  -- we believe, maximally -- simple experimental settings,
for which the first three challenges 
are all comfortably in the realm of current technology.
Our striking underlying technical observation is that 
some of the most interesting collective phenomena predicted for  
$S{=}1/2$ Heisenberg (HB) spin models are in fact not predicated on 
the SU$(2)$ symmetry of Heisenberg models at all. Rather, 
they persist across a wide swathe of exchange anisotropy parameters, 
In cases of interest this includes XY models, which 
lend themselves to cold atomic settings, as the $S{=}1/2$ degree of freedom can 
simply be represented by the presence or absence of a particle, 
requiring  only moderate {\em on-site} interactions to forbid double or higher occupancy. 
This is much easier to generate than interactions between neighbours, which are only expected to
become available once dipolar atoms or molecules get operational for quantum simulations. 
Due to the widespread experimental availability of the bosonic atoms $^{87}$Rb and $^{133}$Cs and the fact that in our 
proposal the spin exchange energy scale is equivalent to the strength of the bosonic hopping matrix 
element, we believe that the energy scales are very favourable for atomic quantum simulations.

The respective spin and bosonic Hamiltonians read 
\begin{eqnarray}
\mathcal{H}_\mathrm{spin}&=&J\sum_{\langle i,j\rangle}  (S^+_i S^-_j +S^-_i S^+_j)/2 + \Delta\ S^z_i S^z_j \\
\mathcal{H}_\mathrm{boson}&=&J\sum_{\langle i,j\rangle}  (b^\dagger_i b^{\phantom{\dagger}}_j +b^{\phantom{\dagger}}_i b^\dagger_j) + \frac{U}{2}\sum_i n_i (n_i-1) \ ,\nonumber
\end{eqnarray}
where $S^z_j$ and $S^\pm_j=S^x_j\pm i S^y_j$ are spin-$1/2$ operators and $b^\dagger_j$ and $n_j=b^\dagger_j b^{\phantom{\dagger}}_j$ 
boson operators at site $j$ of a lattice defined by its bonds $\langle ij \rangle$. We consider antiferromagnetic spin-spin interactions $J>0$
in the following.

The anisotropy parameter $\Delta=1$ for the HB model. Crucially, it vanishes for the XY case, corresponding 
to  $\mathcal{H}_\mathrm{boson}$ in the limit of large $U$. It can thus be realised containing 
a simple hopping term -- that is to say a {\it one-body} 
kinetic term! Thus, the problem 
is reduced to one of engineering an optical lattice for the hardcore bosons -- i.e.\ with only {\em onsite}
interactions --  with $J>0$! 
 
\begin{figure*}
\centerline{\includegraphics*[clip,width=\linewidth]{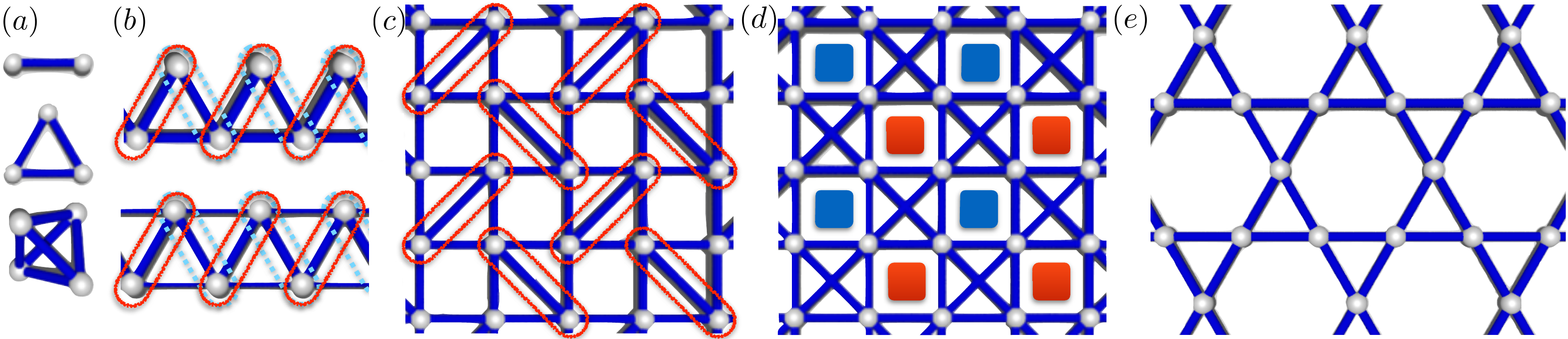}}
\caption{(Color online) Equivalence between ground states of HB and XY models. This holds exactly
for (a) connected clusters such as dimers, 
{{triangles and tetrahedra, as well as (b) the sawtooth and Majumdar-Ghosh chains,
and (c) the Shastry-Sutherland lattice}}, in which spins form singlet pairs (denoted in red). 
It is approximate but accurate for the archetypal frustrated models on 
the well-studied  
{{(d) checkerboard (cf.~Fig.~\ref{fig:checker})  and (e) kagome lattices}}, including the
emergent synthetic gauge field of the former, and the
celebrated spin liquid state of the latter. No simple picture
for the ground state of the kagome magnet is known.}
\label{fig:lattices_equiv}
\end{figure*}

The mapping between hardcore bosons and $S{=}1/2$ XY models 
is well known and the intrinsic interest of hardcore bosonic models with the hopping $J$
as the magnetic energy scale has
also been recognised \cite{Sachdev2002,Simon2011,Eckardt2010,Moller2010,Moller2012,Meinert2014}. Our main 
contribution is to  explicitly demonstrate
that a number of very interesting phenomena do in fact occur in such XY models, 
including the physics of an emergent synthetic gauge field, the realisation
of resonating valence bond physics and the 
appearance of quantum spin liquidity,
topics which have been at the focus of intense attention recently. 

We first present an origin of this remarkable XY-HB `equivalence', and present families of
frustrated lattices where it applies, 
identifying the kagome quantum spin liquid as perhaps
the most exciting and promising target for realisation via quantum simulators. 
We complement this by addressing experimental realisability 
with current AMO technology, and conclude with an outlook.

{\unue{Origin of XY-HB equivalence}} 
Our basic idea can already crisply be gleaned from considering a fully connected cluster of $q$ spins, 
where $q=3 (4)$ corresponds to a triangle (tetrahedron). Up to a constant, the Hamiltonian 
$\mathcal{H}_\mathrm{spin}$ 
for the cluster 
can be written in terms of its total spin, ${\bf L}=\sum_{i=1}^q {\bf S}_i$:
\bea
H_{HB}&=&{\bf L}^2=L(L+1) \\
H_{XY}&=&{\bf L}^2-L_z^2 = L(L+1) - L_z^2\ 
\eea
with $L$ is a (half-)integer between  $L_{min}=0\ (1/2)$ and $L_{max}=q/2$  
for $q$ even (odd), and $-L \leq L_z \leq L$ in integer
steps. 

Thus, {\em for both HB and XY spins} the ground state always has $L=L_z=L_{min}$. 
In both cases -- indeed, for any value of $\Delta$ --  
all states can be completely labelled by assigning values $\{L,L_z\}$ but their
order in energy above the ground state differs between HB and XY -- the equivalence only 
holds at low energy. We believe this is an important ingredient for the quasi-equivalence of 
low-energy states of  XY and HB more generally. 
We next demonstrate the HB-XY equivalence explicitly, for a number of different lattices 
(Fig.~\ref{fig:lattices_equiv}),  among them some of the most studied exotic magnets. 

{\unue{Sawtooth and Majumdar-Ghosh chains}}
The first instances are two triangle-based chains, Fig.~\ref{fig:lattices_equiv}(b), 
where the HB and XY groundstates are exactly identical. 
The ground states are exact valence bond coverings for the HB case,
in which neighbouring spins are paired up into SU$(2)$ singlets, which are also XY ground states.

\unue{Shastry-Sutherland Lattice}
The next exhibit is the Shastry-Sutherland~\cite{Shastry1981b} lattice, Fig.~\ref{fig:lattices_equiv}(c),  
which has received much attention in the context of the material SrCu$_2$(BO$_3$)$_2$~\cite{Kageyama1999}.
Two particular highlights of the Shastry-Sutherland lattice are i) its exact valence bond 
covering ground state for sufficiently strong diagonal
coupling (going back to an exact solution by Shastry and Sutherland~\cite{Shastry1981b}) and ii) a 
fascinatingly rich magnetization plateaux 
structure~\cite{Matsuda2013} exhibiting e.g.~crystals of triplon bound states and even spin supersolids 
\cite{Corboz2014}.

We performed numerical exact diagonalizations of finite XY systems and find that the exact singlet
covering ground state exists when the coupling on the diagonal bonds exceeds $\sim1.6$ times the coupling
on the square lattice bonds. For smaller diagonal couplings a N\'eel state in the spin $x{-}y$ plane is formed. 
This state is equivalent to a staggered superfluid state of the hardcore bosons. Both phases are also 
present in analogous form in the HB phase diagram at zero magnetization~\cite{Lauchli2002,Corboz2013}.

\unue{Checkerboard Lattice} 
The next case, whose analysis is considerably less straightforward, 
is the checkerboard lattice [Fig.~\ref{fig:lattices_equiv}(d)]
also known as the two-dimensional version of the pyrochlore lattice.  Both can be thought of consisting
of  tetrahedra
arranged to share corners, with a tetrahedron projected onto a plane corresponding to a square with
diagonal interactions [Fig.~\ref{fig:lattices_equiv}(a), bottom]. 
The Heisenberg model on the checkerboard lattice 
has been established to exhibit quantum order by disorder, in the
form of a plaquette-type valence bond crystal~\cite{Fouet2003,Moessner2004,Tchernyshyov2003,Shannon2004}. 

\begin{figure}
\includegraphics*[width=\linewidth]{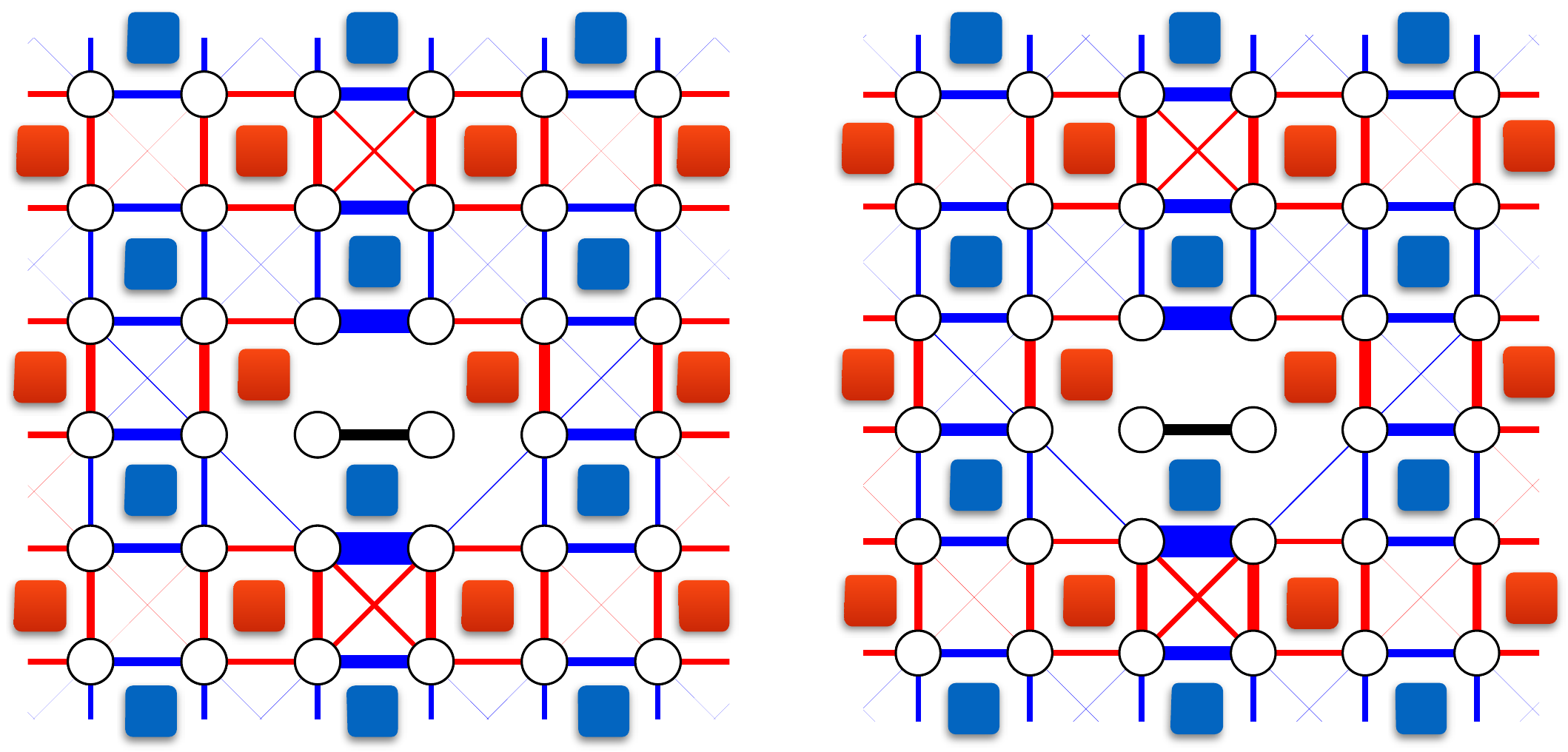}
\caption{(Color online) Valence bond crystal on the checkerboard lattice. 
The two panels show the 
essentially identical pattern
ground state correlations for $N=36$ spins for the XY (left) and HB (right)
models. 
The thickness of the red(blue) lines indicates the degree of suppression(enhancement) of the probability
of finding a valence bond given the presence of the black valence bond, showing the enhancement
on the plaquettes as in Fig~\ref{fig:lattices_equiv}(d).}
\label{fig:checker}
\end{figure}

We have confirmed from finite-size numerical diagonalization studies that for both XY and HB cases,
the same ground-state correlations -- in particular, the same symmetry-broken non-classical 
plaquette valence-bond crystal -- occurs, see Fig.~\ref{fig:checker}. 

\unue{Kagome Lattice}  Perhaps the most surprising
case of our HB-XY equivalence 
is that of the kagome lattice, Fig.~\ref{fig:lattices_equiv}(e). The $S{=}1/2$ HB
antiferromagnet on this lattice has become one of the 
paradigms of the physics of quantum spin liquids. 
Despite intense theoretical efforts for more than twenty years~\cite{Elser1989,Sachdev1992,Lecheminant1997,Mila1998,Ran2007,Yan2011,Depenbrock2012},
further reinforced by the suggestion that 
with Herbertsmithite 
there may or may not be a 
material available which possibly realizes a HB $S{=}1/2$ model with only small deviations~\cite{Shores2005,Mendels2007,Han2012},
 the nature of the ground state and the low lying excitations 
is still not settled. Thus, unlike the previous examples, 
no reliable solution of the kagome $S{=}1/2$ HB model is known, either  
exact or approximate, and it is here that results from 
quantum simulations would be most welcome. 

\begin{figure}[t]
\includegraphics*[width=\linewidth]{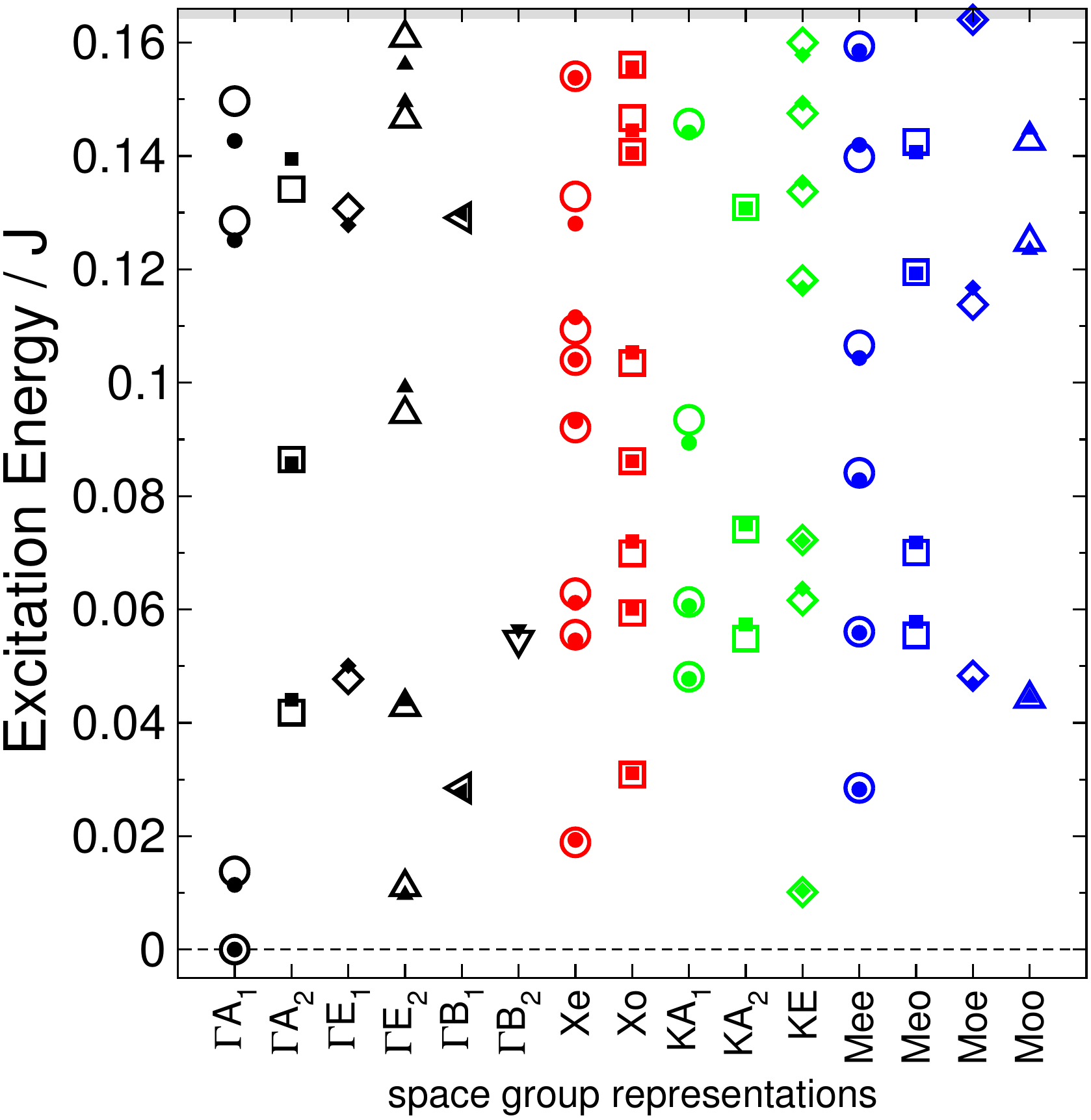}
\caption{(Color online) Kagome lattice: 
The low-lying many body energy spectrum on $N=36$ site sample for HB (open symbols) and 
XY (solid symbols) is identical modulo a single global shift and rescaling.}
\label{fig:kagome_spec}
\end{figure}

The XY-HB equivalence is most starkly illustrated by considering the nature  of the low-energy states
of the two models for a finite-size system.  Compared to  the case of the clusters discussed above,
where the quantum numbers $\{L, L_z\}$ were used to establish a correspondence between the states 
for XY and HB, for a lattice system, the quantum numbers are richer. The states can be grouped into sectors, 
labelled by the 
irreducible representations of the space group of the lattice, with each sector now containing an exponentially large
number of states. The spectrum thus decomposed is shown in 
Fig.~\ref{fig:kagome_spec} for a cluster of 36 sites, where we have 
effected a simple overall rescaling of the energy axis, 
multiplying the excitation energies for  XY compared to HB by a factor of 2.

We thus find that there is a precise pairwise XY-HB correspondence between not only the 
ground~\cite{He2015}
but also {\em each and every} low-energy excited states 
in {\em each and every} sector!  {\em The entire low-energy spectrum is in 
near-perfect correspondence between HB and XY cases!!!} 

\unue{Quantum Simulators}
With this in mind, we now turn to prospects for realising these models in actual cold atomic quantum simulators,
focussing our attention on the cases for which the exact ground states are not analytically available. 
In particular, given the combination of  the great interest in the kagome spin liquid phase and its faithful representation
by an XY  model, we believe that this establishes the hardcore boson model on the kagome lattice alongside
the checkerboard lattice with its emergent gauge field as the most promising target quantum simulator for exotic spin physics. 
As mentioned above, all we need is (a) band structure engineering and (b) hardcore repulsion. 

Regarding (a), optical kagome lattices have
already been obtained and studied~\cite{Jo2012}. The sign of the hopping, 
which needs to be `frustrated' for the spin liquid to appear instead of a conventional superfluid
can be obtained by the by now well-understood protocol involving shaking~\cite{Eckardt2005,Struck2011,Struck2012,Jotzu2014},
therefore changing the signs of all three hopping integrals at once. Another promising setup is to alter the sign of the hopping along only
one of the three linear chain directions of the kagome lattice through Raman-assisted hopping~\cite{Jaksch2003}, as recently experimentally demonstrated
in ladder~\cite{Aidelsburger2011} and square lattice Hofstadter systems~\cite{Aidelsburger2013,Miyake2013}. 
This amounts to a different gauge choice to implement the physical flux $\pi$ (equivalent to the frustrated hopping) through each triangle.
For the checkerboard lattice, we are not aware of any experimental attempts using cold atoms so far,
although promising proposals in this direction making use of the properties of Rydberg atoms have 
been put forward~\cite{Glaetzle2014}. Our proposal supplements these by identifying an avenue requiring no fine-tuning of either interactions or 
hopping -- rather, it suffices to generate a square optical lattice, decorated with the additional checkerboard diagonal hoppings of strength  
in the range $0.75\sim1.25$, according to the HB results~\cite{Sindzingre2002}, without need for particular fine-tuning. 

\begin{figure}
\includegraphics*[width=0.7\linewidth]{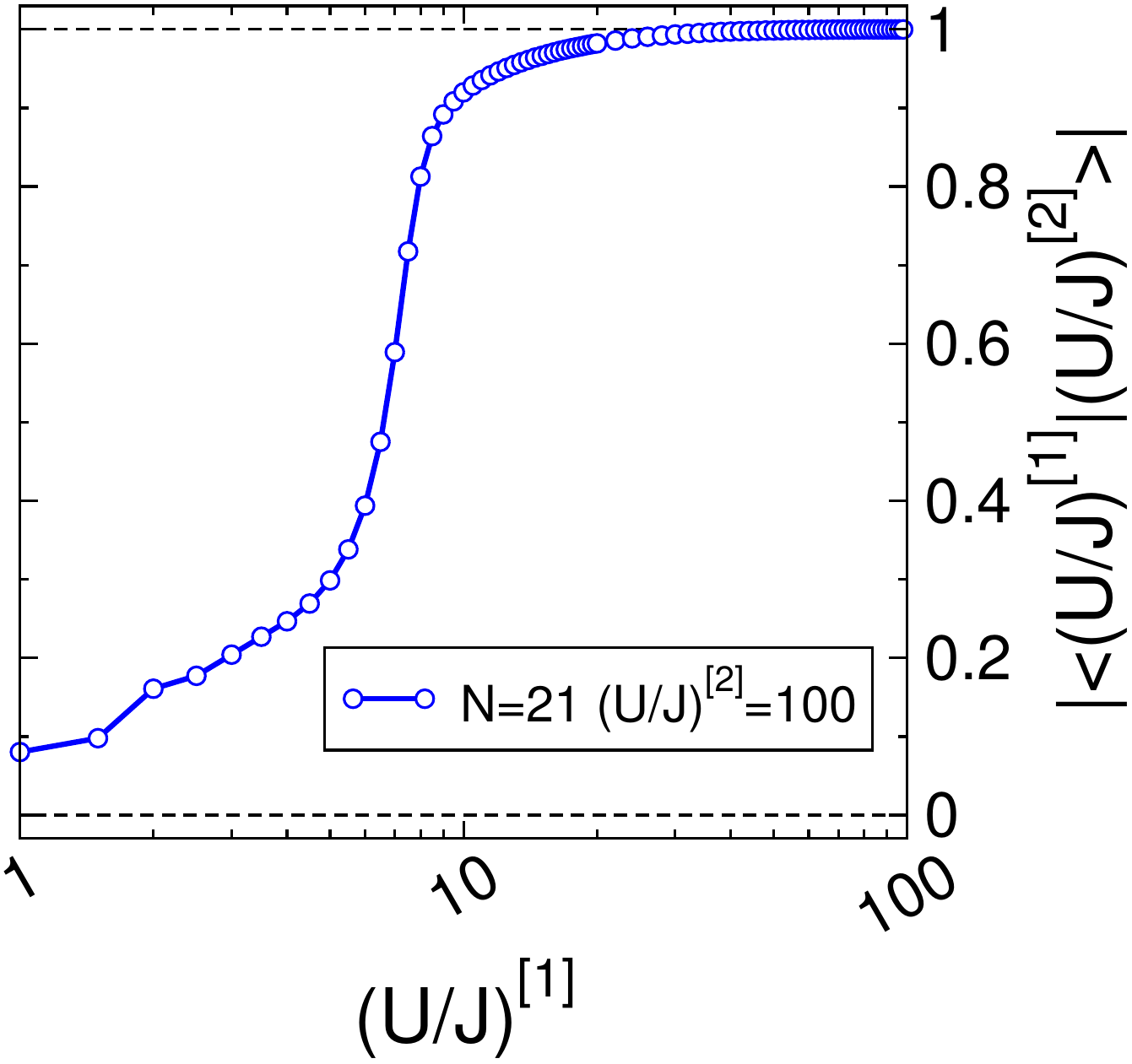}
\caption{(Color online) Kagome lattice: Overlap of the ground state at $U/J=100$ with the ground state at lower $U/J$. 
Down to values of $U/J \sim 10 - 20$ the spin liquid physics will prevail over competing phases.
}
\label{fig:kagome_overlap}
\end{figure}

Regarding (b), we note that it is not even necessary to reach the ideal limit of hardcore bosons 
to access the phases of the XY model. This is true for both checkerboard valence bond crystal~\cite{Glaetzle2014} 
and the kagome spin liquid state, for which (Fig.~\ref{fig:kagome_overlap})
the overlap of the exact XY ground state with that of a model with softened onsite repulsion, $U$, is high for $U/J>10$, 
and almost perfect for $U/J>20$. These conditions are easily met with either $^{87}$Rb or $^{133}$Cs atoms. $^{133}$Cs
has the further advantage of a Feshbach resonance, allowing to access a large $U/J$ ratio while maintaining a sizeable 
hopping energy scale $J$. 

With our proposed kagome quantum simulator one can then first address interesting few particle physics, such as the dynamics of 
caged (localized) magnons~\cite{Schulenburg2002} which could be monitored by single site resolution, then proceed to the 
investigation of spontaneous pattern formation in various magnetisation plateaux~\cite{Nishimoto2013,Capponi2013}, which could be detected 
using Bragg spectroscopy, and then ultimately address the nature of the spin liquid at half filling, where many theoretical questions are still open.

How does our proposal help in terms of 'practicalities' of reaching a regime where the 
cooperative physics of these models is visible? The challenges here 
are very similar for most proposals of atomic quantum 
simulators, with our proposal here perhaps favourable in the following ways. Firstly, we 
invoke exclusively on-site -- and therefore generically strong -- interaction terms rather than 
perturbatively suppressed `second-order' superexchange terms, or weak further-neighbour 
or fine-tuned interactions. Secondly, the other energy scale is given by 
{\em one-body} hopping terms, which can be implemented as `first-order' terms for the kagome lattice, and
which should therefore be comparatively large. Thirdly, since the ratio of hardcore repulsion to hopping
does not need to be all that large,  the different terms in the Hamiltonian do not need span an exorbitant range
of energies by themselves.  Taken together, these provide  a high
level of robustness against coupling to the environment or experimental noise, which
will e.g.\   allow a larger time-window of stability of the system in order to study its real-time dynamics. 
Regarding cooling, while for a strict fixed-entropy preparation protocol, a larger energy scale makes no difference, 
it can be very useful when using a spatially inhomogeneous setting in which another part of the system acts as a `bath'
for the quantum simulator. At any rate, 
a study of the finite-temperature properties of the kagome spin liquid is already be a worthy goal in itself. 

\unue{Summary and outlook}
In summary, we have proposed a set of maximally simple quantum simulators to capture 
the interesting many-body physics of frustrated $S{=}1/2$ HB magnets. This
include various poster children of correlated electron physics and topological phases of 
condensed matter in the form of resonating valence bond phases and
all the way up to the enigmatic kagome spin liquid.
The observation of the persistent plaquette valence-bond crystal in the XY case is all the more intriguing 
when considering that this phase is adiabatically connected to the strong Ising limit, where the XXZ Hamiltonian 
maps onto 
quantum square ice, a quantum link model realising the  $U(1)$ lattice gauge theory (LGT) which has
 an emergent gauge field, albeit confined at zero temperature in $2D$. 
It should thus be feasible to study  LGTs and their emergent  gauge fields
through  implementations that are much simpler than those which attempt to enforce the gauge constraint
in a strict way. 

Many further avenues for research are plainly visible. To start with, 
the  robustness of the kagome spin liquid phase, which persists for a fabulously broad range of transverse 
couplings, obviously requires a deeper understanding.
More broadly, the quantum simulation of XY magnetisation curves -- by tuning the density of bosons -- would link 
up with the rich and as yet not fully understood field  (again, especially for kagome magnets~\cite{Nishimoto2013,Capponi2013}) 
of magnetisation plateau in frustrated quantum magnets.

Moreover, we  also expect an XY-HB 
analogy for a wide range of other lattices which have received considerable interest over the years in 
$d=1,2$ and 3, 
such as the kagome strip~\cite{Waldtmann2000}, the square kagome (squagome) lattice~\cite{Siddharthan2001} and even the hyperkagome lattice~\cite{Okamoto2007}.
For the Husimi cactus \cite{Hao2009}, which has played an important role for understanding
excitations in kagome-type lattice, but is hard to represent as an optical lattice, there is an exact equivalence between 
{\em exponentially many} XY and HB ground states. 

More conceptually, note  that perturbative schemes based on expansions around singlet coverings 
of lattices like the one used to derive the Rokhsar-Kivelson Hamiltonian~\cite{Rokhsar1988} of short-range resonating valence bond 
physics \cite{Moessner2011} will apply not only for HB but also to
 their corresponding XY models. On general grounds, one then expects that {\em gapped} phases, like
a topological $Z_2$ spin liquid~\cite{Moessner2001} with SU(2) symmetry~\cite{Raman2005}, 
will then be stable as one perturbs from the HB towards the XY point.

Beyond this, the potential applicability of these ideas 
for the interesting cases of pyrochlore (the $d=3$ brethren of the checkerboard) or the SCGO lattice 
(a hybrid between pyrochlore and kagome) is entirely open. For these, there are at present {\em no} reliable
analytic or numerical approaches available, and therefore, cold atomic quantum simulators could
establish themselves as the method of choice for studying complex higher-dimensional quantum magnets. 

Overall, exploring the physics of frustrated $S{=}1/2$ XY models via hardcore bosons on optical lattices 
patently has tremendous potential for advancing our
 understanding of, and experimental access to, the physics of quantum spin liquids,
emergent synthetic gauge fields,  and other exotic  phenomena in magnetic quantum matter. 

\section*{Acknowledgements}
We thank N.R. Cooper, C.~Hooley and S.L. Sondhi for useful comments and I.~Bloch,
F.~Meinert and H.-C.~N\"agerl for discussion. 
AML was supported by the FWF (I-1310-N27/DFG FOR1807) and FWF SFB Focus (F-4018-N23)
and HPC resources at MPG RZ Garching, UIBK and VSC2/3.

\bibliographystyle{apsrev4-1}

\bibliography{qs_xyh}

\begin{thebibliography}{53}%
\makeatletter
\providecommand \@ifxundefined [1]{%
 \@ifx{#1\undefined}
}%
\providecommand \@ifnum [1]{%
 \ifnum #1\expandafter \@firstoftwo
 \else \expandafter \@secondoftwo
 \fi
}%
\providecommand \@ifx [1]{%
 \ifx #1\expandafter \@firstoftwo
 \else \expandafter \@secondoftwo
 \fi
}%
\providecommand \natexlab [1]{#1}%
\providecommand \enquote  [1]{``#1''}%
\providecommand \bibnamefont  [1]{#1}%
\providecommand \bibfnamefont [1]{#1}%
\providecommand \citenamefont [1]{#1}%
\providecommand \href@noop [0]{\@secondoftwo}%
\providecommand \href [0]{\begingroup \@sanitize@url \@href}%
\providecommand \@href[1]{\@@startlink{#1}\@@href}%
\providecommand \@@href[1]{\endgroup#1\@@endlink}%
\providecommand \@sanitize@url [0]{\catcode `\\12\catcode `\$12\catcode
  `\&12\catcode `\#12\catcode `\^12\catcode `\_12\catcode `\%12\relax}%
\providecommand \@@startlink[1]{}%
\providecommand \@@endlink[0]{}%
\providecommand \url  [0]{\begingroup\@sanitize@url \@url }%
\providecommand \@url [1]{\endgroup\@href {#1}{\urlprefix }}%
\providecommand \urlprefix  [0]{URL }%
\providecommand \Eprint [0]{\href }%
\providecommand \doibase [0]{http://dx.doi.org/}%
\providecommand \selectlanguage [0]{\@gobble}%
\providecommand \bibinfo  [0]{\@secondoftwo}%
\providecommand \bibfield  [0]{\@secondoftwo}%
\providecommand \translation [1]{[#1]}%
\providecommand \BibitemOpen [0]{}%
\providecommand \bibitemStop [0]{}%
\providecommand \bibitemNoStop [0]{.\EOS\space}%
\providecommand \EOS [0]{\spacefactor3000\relax}%
\providecommand \BibitemShut  [1]{\csname bibitem#1\endcsname}%
\let\auto@bib@innerbib\@empty
\bibitem [{\citenamefont {Anderson}(1987)}]{Anderson1987}%
  \BibitemOpen
  \bibfield  {author} {\bibinfo {author} {\bibfnamefont {P.~W.}\ \bibnamefont
  {Anderson}},\ }\href {\doibase 10.1126/science.235.4793.1196} {\bibfield
  {journal} {\bibinfo  {journal} {Science}\ }\textbf {\bibinfo {volume}
  {235}},\ \bibinfo {pages} {1196} (\bibinfo {year} {1987})}\BibitemShut
  {NoStop}%
\bibitem [{\citenamefont {Kitaev}(2003)}]{Kitaev2003}%
  \BibitemOpen
  \bibfield  {author} {\bibinfo {author} {\bibfnamefont {A.}~\bibnamefont
  {Kitaev}},\ }\href {\doibase http://dx.doi.org/10.1016/S0003-4916(02)00018-0}
  {\bibfield  {journal} {\bibinfo  {journal} {Annals of Physics}\ }\textbf
  {\bibinfo {volume} {303}},\ \bibinfo {pages} {2 } (\bibinfo {year}
  {2003})}\BibitemShut {NoStop}%
\bibitem [{\citenamefont {Kitaev}(2006)}]{Kitaev2006}%
  \BibitemOpen
  \bibfield  {author} {\bibinfo {author} {\bibfnamefont {A.}~\bibnamefont
  {Kitaev}},\ }\href {\doibase http://dx.doi.org/10.1016/j.aop.2005.10.005}
  {\bibfield  {journal} {\bibinfo  {journal} {Annals of Physics}\ }\textbf
  {\bibinfo {volume} {321}},\ \bibinfo {pages} {2 } (\bibinfo {year}
  {2006})}\BibitemShut {NoStop}%
\bibitem [{\citenamefont {Duan}\ \emph {et~al.}(2003)\citenamefont {Duan},
  \citenamefont {Demler},\ and\ \citenamefont {Lukin}}]{Duan2003}%
  \BibitemOpen
  \bibfield  {author} {\bibinfo {author} {\bibfnamefont {L.-M.}\ \bibnamefont
  {Duan}}, \bibinfo {author} {\bibfnamefont {E.}~\bibnamefont {Demler}}, \ and\
  \bibinfo {author} {\bibfnamefont {M.~D.}\ \bibnamefont {Lukin}},\ }\href
  {\doibase 10.1103/PhysRevLett.91.090402} {\bibfield  {journal} {\bibinfo
  {journal} {Phys. Rev. Lett.}\ }\textbf {\bibinfo {volume} {91}},\ \bibinfo
  {pages} {090402} (\bibinfo {year} {2003})}\BibitemShut {NoStop}%
\bibitem [{\citenamefont {Sachdev}\ \emph {et~al.}(2002)\citenamefont
  {Sachdev}, \citenamefont {Sengupta},\ and\ \citenamefont
  {Girvin}}]{Sachdev2002}%
  \BibitemOpen
  \bibfield  {author} {\bibinfo {author} {\bibfnamefont {S.}~\bibnamefont
  {Sachdev}}, \bibinfo {author} {\bibfnamefont {K.}~\bibnamefont {Sengupta}}, \
  and\ \bibinfo {author} {\bibfnamefont {S.~M.}\ \bibnamefont {Girvin}},\
  }\href {\doibase 10.1103/PhysRevB.66.075128} {\bibfield  {journal} {\bibinfo
  {journal} {Phys. Rev. B}\ }\textbf {\bibinfo {volume} {66}},\ \bibinfo
  {pages} {075128} (\bibinfo {year} {2002})}\BibitemShut {NoStop}%
\bibitem [{\citenamefont {Simon}\ \emph {et~al.}(2011)\citenamefont {Simon},
  \citenamefont {Bakr}, \citenamefont {Ma}, \citenamefont {Tai}, \citenamefont
  {Preiss},\ and\ \citenamefont {Greiner}}]{Simon2011}%
  \BibitemOpen
  \bibfield  {author} {\bibinfo {author} {\bibfnamefont {J.}~\bibnamefont
  {Simon}}, \bibinfo {author} {\bibfnamefont {W.~S.}\ \bibnamefont {Bakr}},
  \bibinfo {author} {\bibfnamefont {R.}~\bibnamefont {Ma}}, \bibinfo {author}
  {\bibfnamefont {M.~E.}\ \bibnamefont {Tai}}, \bibinfo {author} {\bibfnamefont
  {P.~M.}\ \bibnamefont {Preiss}}, \ and\ \bibinfo {author} {\bibfnamefont
  {M.}~\bibnamefont {Greiner}},\ }\href {\doibase 10.1038/nature09994}
  {\bibfield  {journal} {\bibinfo  {journal} {Nature}\ }\textbf {\bibinfo
  {volume} {472}},\ \bibinfo {pages} {307} (\bibinfo {year}
  {2011})}\BibitemShut {NoStop}%
\bibitem [{\citenamefont {Eckardt}\ \emph {et~al.}(2010)\citenamefont
  {Eckardt}, \citenamefont {Hauke}, \citenamefont {Soltan-Panahi},
  \citenamefont {Becker}, \citenamefont {Sengstock},\ and\ \citenamefont
  {Lewenstein}}]{Eckardt2010}%
  \BibitemOpen
  \bibfield  {author} {\bibinfo {author} {\bibfnamefont {A.}~\bibnamefont
  {Eckardt}}, \bibinfo {author} {\bibfnamefont {P.}~\bibnamefont {Hauke}},
  \bibinfo {author} {\bibfnamefont {P.}~\bibnamefont {Soltan-Panahi}}, \bibinfo
  {author} {\bibfnamefont {C.}~\bibnamefont {Becker}}, \bibinfo {author}
  {\bibfnamefont {K.}~\bibnamefont {Sengstock}}, \ and\ \bibinfo {author}
  {\bibfnamefont {M.}~\bibnamefont {Lewenstein}},\ }\href
  {http://stacks.iop.org/0295-5075/89/i=1/a=10010} {\bibfield  {journal}
  {\bibinfo  {journal} {EPL (Europhysics Letters)}\ }\textbf {\bibinfo {volume}
  {89}},\ \bibinfo {pages} {10010} (\bibinfo {year} {2010})}\BibitemShut
  {NoStop}%
\bibitem [{\citenamefont {M\"oller}\ and\ \citenamefont
  {Cooper}(2010)}]{Moller2010}%
  \BibitemOpen
  \bibfield  {author} {\bibinfo {author} {\bibfnamefont {G.}~\bibnamefont
  {M\"oller}}\ and\ \bibinfo {author} {\bibfnamefont {N.~R.}\ \bibnamefont
  {Cooper}},\ }\href {\doibase 10.1103/PhysRevA.82.063625} {\bibfield
  {journal} {\bibinfo  {journal} {Phys. Rev. A}\ }\textbf {\bibinfo {volume}
  {82}},\ \bibinfo {pages} {063625} (\bibinfo {year} {2010})}\BibitemShut
  {NoStop}%
\bibitem [{\citenamefont {M\"oller}\ and\ \citenamefont
  {Cooper}(2012)}]{Moller2012}%
  \BibitemOpen
  \bibfield  {author} {\bibinfo {author} {\bibfnamefont {G.}~\bibnamefont
  {M\"oller}}\ and\ \bibinfo {author} {\bibfnamefont {N.~R.}\ \bibnamefont
  {Cooper}},\ }\href {\doibase 10.1103/PhysRevLett.108.045306} {\bibfield
  {journal} {\bibinfo  {journal} {Phys. Rev. Lett.}\ }\textbf {\bibinfo
  {volume} {108}},\ \bibinfo {pages} {045306} (\bibinfo {year}
  {2012})}\BibitemShut {NoStop}%
\bibitem [{\citenamefont {Meinert}\ \emph {et~al.}(2014)\citenamefont
  {Meinert}, \citenamefont {Mark}, \citenamefont {Kirilov}, \citenamefont
  {Lauber}, \citenamefont {Weinmann}, \citenamefont {Gr\"obner}, \citenamefont
  {Daley},\ and\ \citenamefont {N\"agerl}}]{Meinert2014}%
  \BibitemOpen
  \bibfield  {author} {\bibinfo {author} {\bibfnamefont {F.}~\bibnamefont
  {Meinert}}, \bibinfo {author} {\bibfnamefont {M.~J.}\ \bibnamefont {Mark}},
  \bibinfo {author} {\bibfnamefont {E.}~\bibnamefont {Kirilov}}, \bibinfo
  {author} {\bibfnamefont {K.}~\bibnamefont {Lauber}}, \bibinfo {author}
  {\bibfnamefont {P.}~\bibnamefont {Weinmann}}, \bibinfo {author}
  {\bibfnamefont {M.}~\bibnamefont {Gr\"obner}}, \bibinfo {author}
  {\bibfnamefont {A.~J.}\ \bibnamefont {Daley}}, \ and\ \bibinfo {author}
  {\bibfnamefont {H.-C.}\ \bibnamefont {N\"agerl}},\ }\href {\doibase
  10.1126/science.1248402} {\bibfield  {journal} {\bibinfo  {journal}
  {Science}\ }\textbf {\bibinfo {volume} {344}},\ \bibinfo {pages} {1259}
  (\bibinfo {year} {2014})}\BibitemShut {NoStop}%
\bibitem [{\citenamefont {Shastry}\ and\ \citenamefont
  {Sutherland}(1981)}]{Shastry1981b}%
  \BibitemOpen
  \bibfield  {author} {\bibinfo {author} {\bibfnamefont {B.~S.}\ \bibnamefont
  {Shastry}}\ and\ \bibinfo {author} {\bibfnamefont {B.}~\bibnamefont
  {Sutherland}},\ }\href {\doibase
  http://dx.doi.org/10.1016/0378-4363(81)90838-X} {\bibfield  {journal}
  {\bibinfo  {journal} {Physica B+C}\ }\textbf {\bibinfo {volume} {108}},\
  \bibinfo {pages} {1069 } (\bibinfo {year} {1981})}\BibitemShut {NoStop}%
\bibitem [{\citenamefont {Kageyama}\ \emph {et~al.}(1999)\citenamefont
  {Kageyama}, \citenamefont {Yoshimura}, \citenamefont {Stern}, \citenamefont
  {Mushnikov}, \citenamefont {Onizuka}, \citenamefont {Kato}, \citenamefont
  {Kosuge}, \citenamefont {Slichter}, \citenamefont {Goto},\ and\ \citenamefont
  {Ueda}}]{Kageyama1999}%
  \BibitemOpen
  \bibfield  {author} {\bibinfo {author} {\bibfnamefont {H.}~\bibnamefont
  {Kageyama}}, \bibinfo {author} {\bibfnamefont {K.}~\bibnamefont {Yoshimura}},
  \bibinfo {author} {\bibfnamefont {R.}~\bibnamefont {Stern}}, \bibinfo
  {author} {\bibfnamefont {N.~V.}\ \bibnamefont {Mushnikov}}, \bibinfo {author}
  {\bibfnamefont {K.}~\bibnamefont {Onizuka}}, \bibinfo {author} {\bibfnamefont
  {M.}~\bibnamefont {Kato}}, \bibinfo {author} {\bibfnamefont {K.}~\bibnamefont
  {Kosuge}}, \bibinfo {author} {\bibfnamefont {C.~P.}\ \bibnamefont
  {Slichter}}, \bibinfo {author} {\bibfnamefont {T.}~\bibnamefont {Goto}}, \
  and\ \bibinfo {author} {\bibfnamefont {Y.}~\bibnamefont {Ueda}},\ }\href
  {\doibase 10.1103/PhysRevLett.82.3168} {\bibfield  {journal} {\bibinfo
  {journal} {Phys. Rev. Lett.}\ }\textbf {\bibinfo {volume} {82}},\ \bibinfo
  {pages} {3168} (\bibinfo {year} {1999})}\BibitemShut {NoStop}%
\bibitem [{\citenamefont {Matsuda}\ \emph {et~al.}(2013)\citenamefont
  {Matsuda}, \citenamefont {Abe}, \citenamefont {Takeyama}, \citenamefont
  {Kageyama}, \citenamefont {Corboz}, \citenamefont {Honecker}, \citenamefont
  {Manmana}, \citenamefont {Foltin}, \citenamefont {Schmidt},\ and\
  \citenamefont {Mila}}]{Matsuda2013}%
  \BibitemOpen
  \bibfield  {author} {\bibinfo {author} {\bibfnamefont {Y.~H.}\ \bibnamefont
  {Matsuda}}, \bibinfo {author} {\bibfnamefont {N.}~\bibnamefont {Abe}},
  \bibinfo {author} {\bibfnamefont {S.}~\bibnamefont {Takeyama}}, \bibinfo
  {author} {\bibfnamefont {H.}~\bibnamefont {Kageyama}}, \bibinfo {author}
  {\bibfnamefont {P.}~\bibnamefont {Corboz}}, \bibinfo {author} {\bibfnamefont
  {A.}~\bibnamefont {Honecker}}, \bibinfo {author} {\bibfnamefont {S.~R.}\
  \bibnamefont {Manmana}}, \bibinfo {author} {\bibfnamefont {G.~R.}\
  \bibnamefont {Foltin}}, \bibinfo {author} {\bibfnamefont {K.~P.}\
  \bibnamefont {Schmidt}}, \ and\ \bibinfo {author} {\bibfnamefont
  {F.}~\bibnamefont {Mila}},\ }\href {\doibase 10.1103/PhysRevLett.111.137204}
  {\bibfield  {journal} {\bibinfo  {journal} {Phys. Rev. Lett.}\ }\textbf
  {\bibinfo {volume} {111}},\ \bibinfo {pages} {137204} (\bibinfo {year}
  {2013})}\BibitemShut {NoStop}%
\bibitem [{\citenamefont {Corboz}\ and\ \citenamefont
  {Mila}(2014)}]{Corboz2014}%
  \BibitemOpen
  \bibfield  {author} {\bibinfo {author} {\bibfnamefont {P.}~\bibnamefont
  {Corboz}}\ and\ \bibinfo {author} {\bibfnamefont {F.}~\bibnamefont {Mila}},\
  }\href {\doibase 10.1103/PhysRevLett.112.147203} {\bibfield  {journal}
  {\bibinfo  {journal} {Phys. Rev. Lett.}\ }\textbf {\bibinfo {volume} {112}},\
  \bibinfo {pages} {147203} (\bibinfo {year} {2014})}\BibitemShut {NoStop}%
\bibitem [{\citenamefont {L\"auchli}\ \emph {et~al.}(2002)\citenamefont
  {L\"auchli}, \citenamefont {Wessel},\ and\ \citenamefont
  {Sigrist}}]{Lauchli2002}%
  \BibitemOpen
  \bibfield  {author} {\bibinfo {author} {\bibfnamefont {A.}~\bibnamefont
  {L\"auchli}}, \bibinfo {author} {\bibfnamefont {S.}~\bibnamefont {Wessel}}, \
  and\ \bibinfo {author} {\bibfnamefont {M.}~\bibnamefont {Sigrist}},\ }\href
  {\doibase 10.1103/PhysRevB.66.014401} {\bibfield  {journal} {\bibinfo
  {journal} {Phys. Rev. B}\ }\textbf {\bibinfo {volume} {66}},\ \bibinfo
  {pages} {014401} (\bibinfo {year} {2002})}\BibitemShut {NoStop}%
\bibitem [{\citenamefont {Corboz}\ and\ \citenamefont
  {Mila}(2013)}]{Corboz2013}%
  \BibitemOpen
  \bibfield  {author} {\bibinfo {author} {\bibfnamefont {P.}~\bibnamefont
  {Corboz}}\ and\ \bibinfo {author} {\bibfnamefont {F.}~\bibnamefont {Mila}},\
  }\href {\doibase 10.1103/PhysRevB.87.115144} {\bibfield  {journal} {\bibinfo
  {journal} {Phys. Rev. B}\ }\textbf {\bibinfo {volume} {87}},\ \bibinfo
  {pages} {115144} (\bibinfo {year} {2013})}\BibitemShut {NoStop}%
\bibitem [{\citenamefont {Fouet}\ \emph {et~al.}(2003)\citenamefont {Fouet},
  \citenamefont {Mambrini}, \citenamefont {Sindzingre},\ and\ \citenamefont
  {Lhuillier}}]{Fouet2003}%
  \BibitemOpen
  \bibfield  {author} {\bibinfo {author} {\bibfnamefont {J.-B.}\ \bibnamefont
  {Fouet}}, \bibinfo {author} {\bibfnamefont {M.}~\bibnamefont {Mambrini}},
  \bibinfo {author} {\bibfnamefont {P.}~\bibnamefont {Sindzingre}}, \ and\
  \bibinfo {author} {\bibfnamefont {C.}~\bibnamefont {Lhuillier}},\ }\href
  {\doibase 10.1103/PhysRevB.67.054411} {\bibfield  {journal} {\bibinfo
  {journal} {Phys. Rev. B}\ }\textbf {\bibinfo {volume} {67}},\ \bibinfo
  {pages} {054411} (\bibinfo {year} {2003})}\BibitemShut {NoStop}%
\bibitem [{\citenamefont {Moessner}\ \emph {et~al.}(2004)\citenamefont
  {Moessner}, \citenamefont {Tchernyshyov},\ and\ \citenamefont
  {Sondhi}}]{Moessner2004}%
  \BibitemOpen
  \bibfield  {author} {\bibinfo {author} {\bibfnamefont {R.}~\bibnamefont
  {Moessner}}, \bibinfo {author} {\bibfnamefont {O.}~\bibnamefont
  {Tchernyshyov}}, \ and\ \bibinfo {author} {\bibfnamefont {S.}~\bibnamefont
  {Sondhi}},\ }\href {\doibase 10.1023/B:JOSS.0000037247.54022.62} {\bibfield
  {journal} {\bibinfo  {journal} {Journal of Statistical Physics}\ }\textbf
  {\bibinfo {volume} {116}},\ \bibinfo {pages} {755} (\bibinfo {year}
  {2004})}\BibitemShut {NoStop}%
\bibitem [{\citenamefont {Tchernyshyov}\ \emph {et~al.}(2003)\citenamefont
  {Tchernyshyov}, \citenamefont {Starykh}, \citenamefont {Moessner},\ and\
  \citenamefont {Abanov}}]{Tchernyshyov2003}%
  \BibitemOpen
  \bibfield  {author} {\bibinfo {author} {\bibfnamefont {O.}~\bibnamefont
  {Tchernyshyov}}, \bibinfo {author} {\bibfnamefont {O.~A.}\ \bibnamefont
  {Starykh}}, \bibinfo {author} {\bibfnamefont {R.}~\bibnamefont {Moessner}}, \
  and\ \bibinfo {author} {\bibfnamefont {A.~G.}\ \bibnamefont {Abanov}},\
  }\href {\doibase 10.1103/PhysRevB.68.144422} {\bibfield  {journal} {\bibinfo
  {journal} {Phys. Rev. B}\ }\textbf {\bibinfo {volume} {68}},\ \bibinfo
  {pages} {144422} (\bibinfo {year} {2003})}\BibitemShut {NoStop}%
\bibitem [{\citenamefont {Shannon}\ \emph {et~al.}(2004)\citenamefont
  {Shannon}, \citenamefont {Misguich},\ and\ \citenamefont
  {Penc}}]{Shannon2004}%
  \BibitemOpen
  \bibfield  {author} {\bibinfo {author} {\bibfnamefont {N.}~\bibnamefont
  {Shannon}}, \bibinfo {author} {\bibfnamefont {G.}~\bibnamefont {Misguich}}, \
  and\ \bibinfo {author} {\bibfnamefont {K.}~\bibnamefont {Penc}},\ }\href
  {\doibase 10.1103/PhysRevB.69.220403} {\bibfield  {journal} {\bibinfo
  {journal} {Phys. Rev. B}\ }\textbf {\bibinfo {volume} {69}},\ \bibinfo
  {pages} {220403} (\bibinfo {year} {2004})}\BibitemShut {NoStop}%
\bibitem [{\citenamefont {Elser}(1989)}]{Elser1989}%
  \BibitemOpen
  \bibfield  {author} {\bibinfo {author} {\bibfnamefont {V.}~\bibnamefont
  {Elser}},\ }\href {\doibase 10.1103/PhysRevLett.62.2405} {\bibfield
  {journal} {\bibinfo  {journal} {Phys. Rev. Lett.}\ }\textbf {\bibinfo
  {volume} {62}},\ \bibinfo {pages} {2405} (\bibinfo {year}
  {1989})}\BibitemShut {NoStop}%
\bibitem [{\citenamefont {Sachdev}(1992)}]{Sachdev1992}%
  \BibitemOpen
  \bibfield  {author} {\bibinfo {author} {\bibfnamefont {S.}~\bibnamefont
  {Sachdev}},\ }\href {\doibase 10.1103/PhysRevB.45.12377} {\bibfield
  {journal} {\bibinfo  {journal} {Phys. Rev. B}\ }\textbf {\bibinfo {volume}
  {45}},\ \bibinfo {pages} {12377} (\bibinfo {year} {1992})}\BibitemShut
  {NoStop}%
\bibitem [{\citenamefont {Lecheminant}\ \emph {et~al.}(1997)\citenamefont
  {Lecheminant}, \citenamefont {Bernu}, \citenamefont {Lhuillier},
  \citenamefont {Pierre},\ and\ \citenamefont {Sindzingre}}]{Lecheminant1997}%
  \BibitemOpen
  \bibfield  {author} {\bibinfo {author} {\bibfnamefont {P.}~\bibnamefont
  {Lecheminant}}, \bibinfo {author} {\bibfnamefont {B.}~\bibnamefont {Bernu}},
  \bibinfo {author} {\bibfnamefont {C.}~\bibnamefont {Lhuillier}}, \bibinfo
  {author} {\bibfnamefont {L.}~\bibnamefont {Pierre}}, \ and\ \bibinfo {author}
  {\bibfnamefont {P.}~\bibnamefont {Sindzingre}},\ }\href {\doibase
  10.1103/PhysRevB.56.2521} {\bibfield  {journal} {\bibinfo  {journal} {Phys.
  Rev. B}\ }\textbf {\bibinfo {volume} {56}},\ \bibinfo {pages} {2521}
  (\bibinfo {year} {1997})}\BibitemShut {NoStop}%
\bibitem [{\citenamefont {Mila}(1998)}]{Mila1998}%
  \BibitemOpen
  \bibfield  {author} {\bibinfo {author} {\bibfnamefont {F.}~\bibnamefont
  {Mila}},\ }\href {\doibase 10.1103/PhysRevLett.81.2356} {\bibfield  {journal}
  {\bibinfo  {journal} {Phys. Rev. Lett.}\ }\textbf {\bibinfo {volume} {81}},\
  \bibinfo {pages} {2356} (\bibinfo {year} {1998})}\BibitemShut {NoStop}%
\bibitem [{\citenamefont {Ran}\ \emph {et~al.}(2007)\citenamefont {Ran},
  \citenamefont {Hermele}, \citenamefont {Lee},\ and\ \citenamefont
  {Wen}}]{Ran2007}%
  \BibitemOpen
  \bibfield  {author} {\bibinfo {author} {\bibfnamefont {Y.}~\bibnamefont
  {Ran}}, \bibinfo {author} {\bibfnamefont {M.}~\bibnamefont {Hermele}},
  \bibinfo {author} {\bibfnamefont {P.~A.}\ \bibnamefont {Lee}}, \ and\
  \bibinfo {author} {\bibfnamefont {X.-G.}\ \bibnamefont {Wen}},\ }\href
  {\doibase 10.1103/PhysRevLett.98.117205} {\bibfield  {journal} {\bibinfo
  {journal} {Phys. Rev. Lett.}\ }\textbf {\bibinfo {volume} {98}},\ \bibinfo
  {pages} {117205} (\bibinfo {year} {2007})}\BibitemShut {NoStop}%
\bibitem [{\citenamefont {Yan}\ \emph {et~al.}(2011)\citenamefont {Yan},
  \citenamefont {Huse},\ and\ \citenamefont {White}}]{Yan2011}%
  \BibitemOpen
  \bibfield  {author} {\bibinfo {author} {\bibfnamefont {S.}~\bibnamefont
  {Yan}}, \bibinfo {author} {\bibfnamefont {D.~A.}\ \bibnamefont {Huse}}, \
  and\ \bibinfo {author} {\bibfnamefont {S.~R.}\ \bibnamefont {White}},\ }\href
  {\doibase 10.1126/science.1201080} {\bibfield  {journal} {\bibinfo  {journal}
  {Science}\ }\textbf {\bibinfo {volume} {332}},\ \bibinfo {pages} {1173}
  (\bibinfo {year} {2011})}\BibitemShut {NoStop}%
\bibitem [{\citenamefont {Depenbrock}\ \emph {et~al.}(2012)\citenamefont
  {Depenbrock}, \citenamefont {McCulloch},\ and\ \citenamefont
  {Schollw\"ock}}]{Depenbrock2012}%
  \BibitemOpen
  \bibfield  {author} {\bibinfo {author} {\bibfnamefont {S.}~\bibnamefont
  {Depenbrock}}, \bibinfo {author} {\bibfnamefont {I.~P.}\ \bibnamefont
  {McCulloch}}, \ and\ \bibinfo {author} {\bibfnamefont {U.}~\bibnamefont
  {Schollw\"ock}},\ }\href {\doibase 10.1103/PhysRevLett.109.067201} {\bibfield
   {journal} {\bibinfo  {journal} {Phys. Rev. Lett.}\ }\textbf {\bibinfo
  {volume} {109}},\ \bibinfo {pages} {067201} (\bibinfo {year}
  {2012})}\BibitemShut {NoStop}%
\bibitem [{\citenamefont {Shores}\ \emph {et~al.}(2005)\citenamefont {Shores},
  \citenamefont {Nytko}, \citenamefont {Bartlett},\ and\ \citenamefont
  {Nocera}}]{Shores2005}%
  \BibitemOpen
  \bibfield  {author} {\bibinfo {author} {\bibfnamefont {M.~P.}\ \bibnamefont
  {Shores}}, \bibinfo {author} {\bibfnamefont {E.~A.}\ \bibnamefont {Nytko}},
  \bibinfo {author} {\bibfnamefont {B.~M.}\ \bibnamefont {Bartlett}}, \ and\
  \bibinfo {author} {\bibfnamefont {D.~G.}\ \bibnamefont {Nocera}},\ }\href
  {\doibase 10.1021/ja053891p} {\bibfield  {journal} {\bibinfo  {journal}
  {Journal of the American Chemical Society}\ }\textbf {\bibinfo {volume}
  {127}},\ \bibinfo {pages} {13462} (\bibinfo {year} {2005})}\BibitemShut
  {NoStop}%
\bibitem [{\citenamefont {Mendels}\ \emph {et~al.}(2007)\citenamefont
  {Mendels}, \citenamefont {Bert}, \citenamefont {de~Vries}, \citenamefont
  {Olariu}, \citenamefont {Harrison}, \citenamefont {Duc}, \citenamefont
  {Trombe}, \citenamefont {Lord}, \citenamefont {Amato},\ and\ \citenamefont
  {Baines}}]{Mendels2007}%
  \BibitemOpen
  \bibfield  {author} {\bibinfo {author} {\bibfnamefont {P.}~\bibnamefont
  {Mendels}}, \bibinfo {author} {\bibfnamefont {F.}~\bibnamefont {Bert}},
  \bibinfo {author} {\bibfnamefont {M.~A.}\ \bibnamefont {de~Vries}}, \bibinfo
  {author} {\bibfnamefont {A.}~\bibnamefont {Olariu}}, \bibinfo {author}
  {\bibfnamefont {A.}~\bibnamefont {Harrison}}, \bibinfo {author}
  {\bibfnamefont {F.}~\bibnamefont {Duc}}, \bibinfo {author} {\bibfnamefont
  {J.~C.}\ \bibnamefont {Trombe}}, \bibinfo {author} {\bibfnamefont {J.~S.}\
  \bibnamefont {Lord}}, \bibinfo {author} {\bibfnamefont {A.}~\bibnamefont
  {Amato}}, \ and\ \bibinfo {author} {\bibfnamefont {C.}~\bibnamefont
  {Baines}},\ }\href {\doibase 10.1103/PhysRevLett.98.077204} {\bibfield
  {journal} {\bibinfo  {journal} {Phys. Rev. Lett.}\ }\textbf {\bibinfo
  {volume} {98}},\ \bibinfo {pages} {077204} (\bibinfo {year}
  {2007})}\BibitemShut {NoStop}%
\bibitem [{\citenamefont {Han}\ \emph {et~al.}(2012)\citenamefont {Han},
  \citenamefont {Helton}, \citenamefont {Chu}, \citenamefont {Nocera},
  \citenamefont {Rodriguez-Rivera}, \citenamefont {Broholm},\ and\
  \citenamefont {Lee}}]{Han2012}%
  \BibitemOpen
  \bibfield  {author} {\bibinfo {author} {\bibfnamefont {T.-H.}\ \bibnamefont
  {Han}}, \bibinfo {author} {\bibfnamefont {J.~S.}\ \bibnamefont {Helton}},
  \bibinfo {author} {\bibfnamefont {S.}~\bibnamefont {Chu}}, \bibinfo {author}
  {\bibfnamefont {D.~G.}\ \bibnamefont {Nocera}}, \bibinfo {author}
  {\bibfnamefont {J.~A.}\ \bibnamefont {Rodriguez-Rivera}}, \bibinfo {author}
  {\bibfnamefont {C.}~\bibnamefont {Broholm}}, \ and\ \bibinfo {author}
  {\bibfnamefont {Y.~S.}\ \bibnamefont {Lee}},\ }\href {\doibase
  10.1038/nature11659} {\bibfield  {journal} {\bibinfo  {journal} {Nature}\
  }\textbf {\bibinfo {volume} {492}},\ \bibinfo {pages} {406} (\bibinfo {year}
  {2012})}\BibitemShut {NoStop}%
\bibitem [{\citenamefont {He}\ and\ \citenamefont {Chen}(2015)}]{He2015}%
  \BibitemOpen
  \bibfield  {author} {\bibinfo {author} {\bibfnamefont {Y.-C.}\ \bibnamefont
  {He}}\ and\ \bibinfo {author} {\bibfnamefont {Y.}~\bibnamefont {Chen}},\
  }\href {\doibase 10.1103/PhysRevLett.114.037201} {\bibfield  {journal}
  {\bibinfo  {journal} {Phys. Rev. Lett.}\ }\textbf {\bibinfo {volume} {114}},\
  \bibinfo {pages} {037201} (\bibinfo {year} {2015})}\BibitemShut {NoStop}%
\bibitem [{\citenamefont {Jo}\ \emph {et~al.}(2012)\citenamefont {Jo},
  \citenamefont {Guzman}, \citenamefont {Thomas}, \citenamefont {Hosur},
  \citenamefont {Vishwanath},\ and\ \citenamefont {Stamper-Kurn}}]{Jo2012}%
  \BibitemOpen
  \bibfield  {author} {\bibinfo {author} {\bibfnamefont {G.-B.}\ \bibnamefont
  {Jo}}, \bibinfo {author} {\bibfnamefont {J.}~\bibnamefont {Guzman}}, \bibinfo
  {author} {\bibfnamefont {C.~K.}\ \bibnamefont {Thomas}}, \bibinfo {author}
  {\bibfnamefont {P.}~\bibnamefont {Hosur}}, \bibinfo {author} {\bibfnamefont
  {A.}~\bibnamefont {Vishwanath}}, \ and\ \bibinfo {author} {\bibfnamefont
  {D.~M.}\ \bibnamefont {Stamper-Kurn}},\ }\href {\doibase
  10.1103/PhysRevLett.108.045305} {\bibfield  {journal} {\bibinfo  {journal}
  {Phys. Rev. Lett.}\ }\textbf {\bibinfo {volume} {108}},\ \bibinfo {pages}
  {045305} (\bibinfo {year} {2012})}\BibitemShut {NoStop}%
\bibitem [{\citenamefont {Eckardt}\ \emph {et~al.}(2005)\citenamefont
  {Eckardt}, \citenamefont {Weiss},\ and\ \citenamefont
  {Holthaus}}]{Eckardt2005}%
  \BibitemOpen
  \bibfield  {author} {\bibinfo {author} {\bibfnamefont {A.}~\bibnamefont
  {Eckardt}}, \bibinfo {author} {\bibfnamefont {C.}~\bibnamefont {Weiss}}, \
  and\ \bibinfo {author} {\bibfnamefont {M.}~\bibnamefont {Holthaus}},\ }\href
  {\doibase 10.1103/PhysRevLett.95.260404} {\bibfield  {journal} {\bibinfo
  {journal} {Phys. Rev. Lett.}\ }\textbf {\bibinfo {volume} {95}},\ \bibinfo
  {pages} {260404} (\bibinfo {year} {2005})}\BibitemShut {NoStop}%
\bibitem [{\citenamefont {Struck}\ \emph {et~al.}(2011)\citenamefont {Struck},
  \citenamefont {\"Olschl\"ager}, \citenamefont {Le~Targat}, \citenamefont
  {Soltan-Panahi}, \citenamefont {Eckardt}, \citenamefont {Lewenstein},
  \citenamefont {Windpassinger},\ and\ \citenamefont {Sengstock}}]{Struck2011}%
  \BibitemOpen
  \bibfield  {author} {\bibinfo {author} {\bibfnamefont {J.}~\bibnamefont
  {Struck}}, \bibinfo {author} {\bibfnamefont {C.}~\bibnamefont
  {\"Olschl\"ager}}, \bibinfo {author} {\bibfnamefont {R.}~\bibnamefont
  {Le~Targat}}, \bibinfo {author} {\bibfnamefont {P.}~\bibnamefont
  {Soltan-Panahi}}, \bibinfo {author} {\bibfnamefont {A.}~\bibnamefont
  {Eckardt}}, \bibinfo {author} {\bibfnamefont {M.}~\bibnamefont {Lewenstein}},
  \bibinfo {author} {\bibfnamefont {P.}~\bibnamefont {Windpassinger}}, \ and\
  \bibinfo {author} {\bibfnamefont {K.}~\bibnamefont {Sengstock}},\ }\href
  {\doibase 10.1126/science.1207239} {\bibfield  {journal} {\bibinfo  {journal}
  {Science}\ }\textbf {\bibinfo {volume} {333}},\ \bibinfo {pages} {996}
  (\bibinfo {year} {2011})}\BibitemShut {NoStop}%
\bibitem [{\citenamefont {Struck}\ \emph {et~al.}(2012)\citenamefont {Struck},
  \citenamefont {\"Olschl\"ager}, \citenamefont {Weinberg}, \citenamefont
  {Hauke}, \citenamefont {Simonet}, \citenamefont {Eckardt}, \citenamefont
  {Lewenstein}, \citenamefont {Sengstock},\ and\ \citenamefont
  {Windpassinger}}]{Struck2012}%
  \BibitemOpen
  \bibfield  {author} {\bibinfo {author} {\bibfnamefont {J.}~\bibnamefont
  {Struck}}, \bibinfo {author} {\bibfnamefont {C.}~\bibnamefont
  {\"Olschl\"ager}}, \bibinfo {author} {\bibfnamefont {M.}~\bibnamefont
  {Weinberg}}, \bibinfo {author} {\bibfnamefont {P.}~\bibnamefont {Hauke}},
  \bibinfo {author} {\bibfnamefont {J.}~\bibnamefont {Simonet}}, \bibinfo
  {author} {\bibfnamefont {A.}~\bibnamefont {Eckardt}}, \bibinfo {author}
  {\bibfnamefont {M.}~\bibnamefont {Lewenstein}}, \bibinfo {author}
  {\bibfnamefont {K.}~\bibnamefont {Sengstock}}, \ and\ \bibinfo {author}
  {\bibfnamefont {P.}~\bibnamefont {Windpassinger}},\ }\href {\doibase
  10.1103/PhysRevLett.108.225304} {\bibfield  {journal} {\bibinfo  {journal}
  {Phys. Rev. Lett.}\ }\textbf {\bibinfo {volume} {108}},\ \bibinfo {pages}
  {225304} (\bibinfo {year} {2012})}\BibitemShut {NoStop}%
\bibitem [{\citenamefont {Jotzu}\ \emph {et~al.}(2014)\citenamefont {Jotzu},
  \citenamefont {Messer}, \citenamefont {Desbuquois}, \citenamefont {Lebrat},
  \citenamefont {Uehlinger}, \citenamefont {Greif},\ and\ \citenamefont
  {Esslinger}}]{Jotzu2014}%
  \BibitemOpen
  \bibfield  {author} {\bibinfo {author} {\bibfnamefont {G.}~\bibnamefont
  {Jotzu}}, \bibinfo {author} {\bibfnamefont {M.}~\bibnamefont {Messer}},
  \bibinfo {author} {\bibfnamefont {R.}~\bibnamefont {Desbuquois}}, \bibinfo
  {author} {\bibfnamefont {M.}~\bibnamefont {Lebrat}}, \bibinfo {author}
  {\bibfnamefont {T.}~\bibnamefont {Uehlinger}}, \bibinfo {author}
  {\bibfnamefont {D.}~\bibnamefont {Greif}}, \ and\ \bibinfo {author}
  {\bibfnamefont {T.}~\bibnamefont {Esslinger}},\ }\href {\doibase
  10.1038/nature13915} {\bibfield  {journal} {\bibinfo  {journal} {Nature}\
  }\textbf {\bibinfo {volume} {515}},\ \bibinfo {pages} {237} (\bibinfo {year}
  {2014})}\BibitemShut {NoStop}%
\bibitem [{\citenamefont {Jaksch}\ and\ \citenamefont
  {Zoller}(2003)}]{Jaksch2003}%
  \BibitemOpen
  \bibfield  {author} {\bibinfo {author} {\bibfnamefont {D.}~\bibnamefont
  {Jaksch}}\ and\ \bibinfo {author} {\bibfnamefont {P.}~\bibnamefont
  {Zoller}},\ }\href {http://stacks.iop.org/1367-2630/5/i=1/a=356} {\bibfield
  {journal} {\bibinfo  {journal} {New Journal of Physics}\ }\textbf {\bibinfo
  {volume} {5}},\ \bibinfo {pages} {56} (\bibinfo {year} {2003})}\BibitemShut
  {NoStop}%
\bibitem [{\citenamefont {Aidelsburger}\ \emph {et~al.}(2011)\citenamefont
  {Aidelsburger}, \citenamefont {Atala}, \citenamefont {Nascimb\`ene},
  \citenamefont {Trotzky}, \citenamefont {Chen},\ and\ \citenamefont
  {Bloch}}]{Aidelsburger2011}%
  \BibitemOpen
  \bibfield  {author} {\bibinfo {author} {\bibfnamefont {M.}~\bibnamefont
  {Aidelsburger}}, \bibinfo {author} {\bibfnamefont {M.}~\bibnamefont {Atala}},
  \bibinfo {author} {\bibfnamefont {S.}~\bibnamefont {Nascimb\`ene}}, \bibinfo
  {author} {\bibfnamefont {S.}~\bibnamefont {Trotzky}}, \bibinfo {author}
  {\bibfnamefont {Y.-A.}\ \bibnamefont {Chen}}, \ and\ \bibinfo {author}
  {\bibfnamefont {I.}~\bibnamefont {Bloch}},\ }\href {\doibase
  10.1103/PhysRevLett.107.255301} {\bibfield  {journal} {\bibinfo  {journal}
  {Phys. Rev. Lett.}\ }\textbf {\bibinfo {volume} {107}},\ \bibinfo {pages}
  {255301} (\bibinfo {year} {2011})}\BibitemShut {NoStop}%
\bibitem [{\citenamefont {Aidelsburger}\ \emph {et~al.}(2013)\citenamefont
  {Aidelsburger}, \citenamefont {Atala}, \citenamefont {Lohse}, \citenamefont
  {Barreiro}, \citenamefont {Paredes},\ and\ \citenamefont
  {Bloch}}]{Aidelsburger2013}%
  \BibitemOpen
  \bibfield  {author} {\bibinfo {author} {\bibfnamefont {M.}~\bibnamefont
  {Aidelsburger}}, \bibinfo {author} {\bibfnamefont {M.}~\bibnamefont {Atala}},
  \bibinfo {author} {\bibfnamefont {M.}~\bibnamefont {Lohse}}, \bibinfo
  {author} {\bibfnamefont {J.~T.}\ \bibnamefont {Barreiro}}, \bibinfo {author}
  {\bibfnamefont {B.}~\bibnamefont {Paredes}}, \ and\ \bibinfo {author}
  {\bibfnamefont {I.}~\bibnamefont {Bloch}},\ }\href {\doibase
  10.1103/PhysRevLett.111.185301} {\bibfield  {journal} {\bibinfo  {journal}
  {Phys. Rev. Lett.}\ }\textbf {\bibinfo {volume} {111}},\ \bibinfo {pages}
  {185301} (\bibinfo {year} {2013})}\BibitemShut {NoStop}%
\bibitem [{\citenamefont {Miyake}\ \emph {et~al.}(2013)\citenamefont {Miyake},
  \citenamefont {Siviloglou}, \citenamefont {Kennedy}, \citenamefont {Burton},\
  and\ \citenamefont {Ketterle}}]{Miyake2013}%
  \BibitemOpen
  \bibfield  {author} {\bibinfo {author} {\bibfnamefont {H.}~\bibnamefont
  {Miyake}}, \bibinfo {author} {\bibfnamefont {G.~A.}\ \bibnamefont
  {Siviloglou}}, \bibinfo {author} {\bibfnamefont {C.~J.}\ \bibnamefont
  {Kennedy}}, \bibinfo {author} {\bibfnamefont {W.~C.}\ \bibnamefont {Burton}},
  \ and\ \bibinfo {author} {\bibfnamefont {W.}~\bibnamefont {Ketterle}},\
  }\href {\doibase 10.1103/PhysRevLett.111.185302} {\bibfield  {journal}
  {\bibinfo  {journal} {Phys. Rev. Lett.}\ }\textbf {\bibinfo {volume} {111}},\
  \bibinfo {pages} {185302} (\bibinfo {year} {2013})}\BibitemShut {NoStop}%
\bibitem [{\citenamefont {Glaetzle}\ \emph {et~al.}(2014)\citenamefont
  {Glaetzle}, \citenamefont {Dalmonte}, \citenamefont {Nath}, \citenamefont
  {Rousochatzakis}, \citenamefont {Moessner},\ and\ \citenamefont
  {Zoller}}]{Glaetzle2014}%
  \BibitemOpen
  \bibfield  {author} {\bibinfo {author} {\bibfnamefont {A.~W.}\ \bibnamefont
  {Glaetzle}}, \bibinfo {author} {\bibfnamefont {M.}~\bibnamefont {Dalmonte}},
  \bibinfo {author} {\bibfnamefont {R.}~\bibnamefont {Nath}}, \bibinfo {author}
  {\bibfnamefont {I.}~\bibnamefont {Rousochatzakis}}, \bibinfo {author}
  {\bibfnamefont {R.}~\bibnamefont {Moessner}}, \ and\ \bibinfo {author}
  {\bibfnamefont {P.}~\bibnamefont {Zoller}},\ }\href {\doibase
  10.1103/PhysRevX.4.041037} {\bibfield  {journal} {\bibinfo  {journal} {Phys.
  Rev. X}\ }\textbf {\bibinfo {volume} {4}},\ \bibinfo {pages} {041037}
  (\bibinfo {year} {2014})}\BibitemShut {NoStop}%
\bibitem [{\citenamefont {Sindzingre}\ \emph {et~al.}(2002)\citenamefont
  {Sindzingre}, \citenamefont {Fouet},\ and\ \citenamefont
  {Lhuillier}}]{Sindzingre2002}%
  \BibitemOpen
  \bibfield  {author} {\bibinfo {author} {\bibfnamefont {P.}~\bibnamefont
  {Sindzingre}}, \bibinfo {author} {\bibfnamefont {J.-B.}\ \bibnamefont
  {Fouet}}, \ and\ \bibinfo {author} {\bibfnamefont {C.}~\bibnamefont
  {Lhuillier}},\ }\href {\doibase 10.1103/PhysRevB.66.174424} {\bibfield
  {journal} {\bibinfo  {journal} {Phys. Rev. B}\ }\textbf {\bibinfo {volume}
  {66}},\ \bibinfo {pages} {174424} (\bibinfo {year} {2002})}\BibitemShut
  {NoStop}%
\bibitem [{\citenamefont {Schulenburg}\ \emph {et~al.}(2002)\citenamefont
  {Schulenburg}, \citenamefont {Honecker}, \citenamefont {Schnack},
  \citenamefont {Richter},\ and\ \citenamefont {Schmidt}}]{Schulenburg2002}%
  \BibitemOpen
  \bibfield  {author} {\bibinfo {author} {\bibfnamefont {J.}~\bibnamefont
  {Schulenburg}}, \bibinfo {author} {\bibfnamefont {A.}~\bibnamefont
  {Honecker}}, \bibinfo {author} {\bibfnamefont {J.}~\bibnamefont {Schnack}},
  \bibinfo {author} {\bibfnamefont {J.}~\bibnamefont {Richter}}, \ and\
  \bibinfo {author} {\bibfnamefont {H.-J.}\ \bibnamefont {Schmidt}},\ }\href
  {\doibase 10.1103/PhysRevLett.88.167207} {\bibfield  {journal} {\bibinfo
  {journal} {Phys. Rev. Lett.}\ }\textbf {\bibinfo {volume} {88}},\ \bibinfo
  {pages} {167207} (\bibinfo {year} {2002})}\BibitemShut {NoStop}%
\bibitem [{\citenamefont {Nishimoto}\ \emph {et~al.}(2013)\citenamefont
  {Nishimoto}, \citenamefont {Shibata},\ and\ \citenamefont
  {Hotta}}]{Nishimoto2013}%
  \BibitemOpen
  \bibfield  {author} {\bibinfo {author} {\bibfnamefont {S.}~\bibnamefont
  {Nishimoto}}, \bibinfo {author} {\bibfnamefont {N.}~\bibnamefont {Shibata}},
  \ and\ \bibinfo {author} {\bibfnamefont {C.}~\bibnamefont {Hotta}},\ }\href
  {http://dx.doi.org/10.1038/ncomms3287} {\bibfield  {journal} {\bibinfo
  {journal} {Nat Commun}\ }\textbf {\bibinfo {volume} {4}} (\bibinfo {year}
  {2013})}\BibitemShut {NoStop}%
\bibitem [{\citenamefont {Capponi}\ \emph {et~al.}(2013)\citenamefont
  {Capponi}, \citenamefont {Derzhko}, \citenamefont {Honecker}, \citenamefont
  {L\"auchli},\ and\ \citenamefont {Richter}}]{Capponi2013}%
  \BibitemOpen
  \bibfield  {author} {\bibinfo {author} {\bibfnamefont {S.}~\bibnamefont
  {Capponi}}, \bibinfo {author} {\bibfnamefont {O.}~\bibnamefont {Derzhko}},
  \bibinfo {author} {\bibfnamefont {A.}~\bibnamefont {Honecker}}, \bibinfo
  {author} {\bibfnamefont {A.~M.}\ \bibnamefont {L\"auchli}}, \ and\ \bibinfo
  {author} {\bibfnamefont {J.}~\bibnamefont {Richter}},\ }\href {\doibase
  10.1103/PhysRevB.88.144416} {\bibfield  {journal} {\bibinfo  {journal} {Phys.
  Rev. B}\ }\textbf {\bibinfo {volume} {88}},\ \bibinfo {pages} {144416}
  (\bibinfo {year} {2013})}\BibitemShut {NoStop}%
\bibitem [{\citenamefont {Waldtmann}\ \emph {et~al.}(2000)\citenamefont
  {Waldtmann}, \citenamefont {Kreutzmann}, \citenamefont {Schollw\"ock},
  \citenamefont {Maisinger},\ and\ \citenamefont {Everts}}]{Waldtmann2000}%
  \BibitemOpen
  \bibfield  {author} {\bibinfo {author} {\bibfnamefont {C.}~\bibnamefont
  {Waldtmann}}, \bibinfo {author} {\bibfnamefont {H.}~\bibnamefont
  {Kreutzmann}}, \bibinfo {author} {\bibfnamefont {U.}~\bibnamefont
  {Schollw\"ock}}, \bibinfo {author} {\bibfnamefont {K.}~\bibnamefont
  {Maisinger}}, \ and\ \bibinfo {author} {\bibfnamefont {H.-U.}\ \bibnamefont
  {Everts}},\ }\href {\doibase 10.1103/PhysRevB.62.9472} {\bibfield  {journal}
  {\bibinfo  {journal} {Phys. Rev. B}\ }\textbf {\bibinfo {volume} {62}},\
  \bibinfo {pages} {9472} (\bibinfo {year} {2000})}\BibitemShut {NoStop}%
\bibitem [{\citenamefont {Siddharthan}\ and\ \citenamefont
  {Georges}(2001)}]{Siddharthan2001}%
  \BibitemOpen
  \bibfield  {author} {\bibinfo {author} {\bibfnamefont {R.}~\bibnamefont
  {Siddharthan}}\ and\ \bibinfo {author} {\bibfnamefont {A.}~\bibnamefont
  {Georges}},\ }\href {\doibase 10.1103/PhysRevB.65.014417} {\bibfield
  {journal} {\bibinfo  {journal} {Phys. Rev. B}\ }\textbf {\bibinfo {volume}
  {65}},\ \bibinfo {pages} {014417} (\bibinfo {year} {2001})}\BibitemShut
  {NoStop}%
\bibitem [{\citenamefont {Okamoto}\ \emph {et~al.}(2007)\citenamefont
  {Okamoto}, \citenamefont {Nohara}, \citenamefont {Aruga-Katori},\ and\
  \citenamefont {Takagi}}]{Okamoto2007}%
  \BibitemOpen
  \bibfield  {author} {\bibinfo {author} {\bibfnamefont {Y.}~\bibnamefont
  {Okamoto}}, \bibinfo {author} {\bibfnamefont {M.}~\bibnamefont {Nohara}},
  \bibinfo {author} {\bibfnamefont {H.}~\bibnamefont {Aruga-Katori}}, \ and\
  \bibinfo {author} {\bibfnamefont {H.}~\bibnamefont {Takagi}},\ }\href
  {\doibase 10.1103/PhysRevLett.99.137207} {\bibfield  {journal} {\bibinfo
  {journal} {Phys. Rev. Lett.}\ }\textbf {\bibinfo {volume} {99}},\ \bibinfo
  {pages} {137207} (\bibinfo {year} {2007})}\BibitemShut {NoStop}%
\bibitem [{\citenamefont {Hao}\ and\ \citenamefont
  {Tchernyshyov}(2009)}]{Hao2009}%
  \BibitemOpen
  \bibfield  {author} {\bibinfo {author} {\bibfnamefont {Z.}~\bibnamefont
  {Hao}}\ and\ \bibinfo {author} {\bibfnamefont {O.}~\bibnamefont
  {Tchernyshyov}},\ }\href {\doibase 10.1103/PhysRevLett.103.187203} {\bibfield
   {journal} {\bibinfo  {journal} {Phys. Rev. Lett.}\ }\textbf {\bibinfo
  {volume} {103}},\ \bibinfo {pages} {187203} (\bibinfo {year}
  {2009})}\BibitemShut {NoStop}%
\bibitem [{\citenamefont {Rokhsar}\ and\ \citenamefont
  {Kivelson}(1988)}]{Rokhsar1988}%
  \BibitemOpen
  \bibfield  {author} {\bibinfo {author} {\bibfnamefont {D.~S.}\ \bibnamefont
  {Rokhsar}}\ and\ \bibinfo {author} {\bibfnamefont {S.~A.}\ \bibnamefont
  {Kivelson}},\ }\href {\doibase 10.1103/PhysRevLett.61.2376} {\bibfield
  {journal} {\bibinfo  {journal} {Phys. Rev. Lett.}\ }\textbf {\bibinfo
  {volume} {61}},\ \bibinfo {pages} {2376} (\bibinfo {year}
  {1988})}\BibitemShut {NoStop}%
\bibitem [{\citenamefont {Moessner}\ and\ \citenamefont
  {Raman}(2011)}]{Moessner2011}%
  \BibitemOpen
  \bibfield  {author} {\bibinfo {author} {\bibfnamefont {R.}~\bibnamefont
  {Moessner}}\ and\ \bibinfo {author} {\bibfnamefont {K.}~\bibnamefont
  {Raman}},\ }in\ \href {\doibase 10.1007/978-3-642-10589-0_17} {\emph
  {\bibinfo {booktitle} {Introduction to Frustrated Magnetism}}},\ \bibinfo
  {series} {Springer Series in Solid-State Sciences}, Vol.\ \bibinfo {volume}
  {164},\ \bibinfo {editor} {edited by\ \bibinfo {editor} {\bibfnamefont
  {C.}~\bibnamefont {Lacroix}}, \bibinfo {editor} {\bibfnamefont
  {P.}~\bibnamefont {Mendels}}, \ and\ \bibinfo {editor} {\bibfnamefont
  {F.}~\bibnamefont {Mila}}}\ (\bibinfo  {publisher} {Springer Berlin
  Heidelberg},\ \bibinfo {year} {2011})\ pp.\ \bibinfo {pages}
  {437--479}\BibitemShut {NoStop}%
\bibitem [{\citenamefont {Moessner}\ and\ \citenamefont
  {Sondhi}(2001)}]{Moessner2001}%
  \BibitemOpen
  \bibfield  {author} {\bibinfo {author} {\bibfnamefont {R.}~\bibnamefont
  {Moessner}}\ and\ \bibinfo {author} {\bibfnamefont {S.~L.}\ \bibnamefont
  {Sondhi}},\ }\href {\doibase 10.1103/PhysRevLett.86.1881} {\bibfield
  {journal} {\bibinfo  {journal} {Phys. Rev. Lett.}\ }\textbf {\bibinfo
  {volume} {86}},\ \bibinfo {pages} {1881} (\bibinfo {year}
  {2001})}\BibitemShut {NoStop}%
\bibitem [{\citenamefont {Raman}\ \emph {et~al.}(2005)\citenamefont {Raman},
  \citenamefont {Moessner},\ and\ \citenamefont {Sondhi}}]{Raman2005}%
  \BibitemOpen
  \bibfield  {author} {\bibinfo {author} {\bibfnamefont {K.~S.}\ \bibnamefont
  {Raman}}, \bibinfo {author} {\bibfnamefont {R.}~\bibnamefont {Moessner}}, \
  and\ \bibinfo {author} {\bibfnamefont {S.~L.}\ \bibnamefont {Sondhi}},\
  }\href {\doibase 10.1103/PhysRevB.72.064413} {\bibfield  {journal} {\bibinfo
  {journal} {Phys. Rev. B}\ }\textbf {\bibinfo {volume} {72}},\ \bibinfo
  {pages} {064413} (\bibinfo {year} {2005})}\BibitemShut {NoStop}%
\end{thebibliography}%

\end{document}